\documentclass[preprint,flushrt]{aastex}
\usepackage{graphicx,amssymb,amsmath,times}
\usepackage{natbib}
\usepackage{lineno}
\newcommand{\MeV}{\mbox{\ensuremath{\mathrm{~MeV}}}}
\newcommand{\GeV}{\mbox{\ensuremath{\mathrm{~GeV}}}}

\def\tc{3C~273}
\def\tct{3C~273~}

\def\fluence{ph cm$^{-2}$}
\def\latflux{ph cm$^{-2}$ s$^{-1}$}
\def\latfluxv{$\times 10^{-5}$ ph cm$^{-2}$ s$^{-1}$}
\def\latfluxvi{$\times 10^{-6}$ ph cm$^{-2}$ s$^{-1}$}

\def\latfluxviiii{$\times 10^{-8}$ ph cm$^{-2}$ s$^{-1}$ h$^{-1}$}
\def\latfluxviiiii{$\times 10^{-7}$ ph cm$^{-2}$ s$^{-1}$}

\linenumbers

\begin{document}
%
%
%
\title{{\it Fermi}-LAT observations of the  exceptional gamma-ray outbursts of 3C~273 in September 2009}

\slugcomment{To be submitted to Astrophysical Journal Letters-}

%

\shorttitle{The exceptional gamma-ray outbursts of 3C~273}
\shortauthors{Abdo et al.}

%
%
\author{
A.~A.~Abdo\altaffilmark{1,2},
M.~Ackermann\altaffilmark{3},
M.~Ajello\altaffilmark{3},
L.~Baldini\altaffilmark{4},
J.~Ballet\altaffilmark{5},
G.~Barbiellini\altaffilmark{6,7},
D.~Bastieri\altaffilmark{8,9},
K.~Bechtol\altaffilmark{3},
R.~Bellazzini\altaffilmark{4},
B.~Berenji\altaffilmark{3}
% ...
}
% ...
\altaffiltext{1}{Space Science Division, Naval Research Laboratory, Washington, DC 20375, USA}
\altaffiltext{2}{National Research Council Research Associate, National Academy of Sciences, Washington, DC 20001, USA}
\altaffiltext{3}{W. W. Hansen Experimental Physics Laboratory, Kavli Institute for Particle Astrophysics and Cosmology, Department of Physics and SLAC National Accelerator Laboratory, Stanford University, Stanford, CA 94305, USA}
\altaffiltext{4}{Istituto Nazionale di Fisica Nucleare, Sezione di Pisa, I-56127 Pisa, Italy}
\altaffiltext{5}{Laboratoire AIM, CEA-IRFU/CNRS/Universit\'e Paris Diderot, Service d'Astrophysique, CEA Saclay, 91191 Gif sur Yvette, France}
\altaffiltext{6}{Istituto Nazionale di Fisica Nucleare, Sezione di Trieste, I-34127 Trieste, Italy}
\altaffiltext{7}{Dipartimento di Fisica, Universit\`a di Trieste, I-34127 Trieste, Italy}
\altaffiltext{8}{Istituto Nazionale di Fisica Nucleare, Sezione di Padova, I-35131 Padova, Italy}
\altaffiltext{9}{Dipartimento di Fisica ``G. Galilei", Universit\`a di Padova, I-35131 Padova, Italy}
% ...

\begin{abstract}
%
We present the light curves and spectral data of two exceptionally luminous
gamma-ray outburts observed by the Large Area Telescope (LAT) experiment on board {\it Fermi} Gamma-ray Space Telescope from 3C~273 in September 2009.
During these flares, having a duration of a few days, the source  reached its highest $\gamma$-ray flux ever measured. This allowed us to study in some details their spectral and temporal structures. The rise and decay are asymmetric on timescales of 6 hours, and  the spectral index was significantly harder during the flares than during the preceding 11 months. We also found that short, very intense flares put out the same time-integrated energy as long, less intense flares like that observed in August 2009.

\end{abstract}



\section{Introduction}
The LAT experiment, on board the {\it Fermi Gamma-ray Space Telescope} satellite,
observes the entire sky, in the 0.02 -- $>$300 \GeV~ band, once every $\sim$3 hours.
It is providing the first collection of well sampled gamma-ray light curves
of several blazars useful to study their  variability on time scales from day to several months (see e.g. Abdo et al. 2010c). Daily light curves can be obtained for several blazars and for those exceptionally bright it is possible to observe significant occasional variations on timescales shorter than a day.  We have now observed such events in 3C273, the nearest quasar.


\tc, the first quasar discovered by Schimdt (1963) and the
first extragalactic source detected by {\it COS}-B in the gamma-ray band (Swanenburg
et al. 1978), is one of the most extensively studied AGN across the entire
electromagnetic spectrum.
It is classified as a flat spectrum radio quasar (FSRQ) and has a redshift $z=0.158$
(see e.g., Strauss et al., 1992).
It was observed by EGRET (3EG~J1229$+$0210 in Hartman et al. 1999) at an average flux
of $0.18$ \latfluxvi with a peak flux of $1.27$\latfluxvi (Nandikotkur et al. 2007).
Despite the huge amount of data collected to now (see for instance Soldi et al. 2008),
its behaviour is still surprising and raises new challenges for physical models. It  was detected by the LAT experiment since the beginning of the observation in 2008
(Marelli, 2008) and is identified with the $gamma$-ray source 1FGL~J1229.1$+$0203 in the First {\it Fermi} LAT catalog (Abdo et al. 2010a). For about one year its behavior was characterized
by a rather stable flux of $\sim$0.3\latfluxvi, with some flares superposed reaching
peak level higher by about an order of magnitude.

Beginning at the end of July 2009, \tct started a brightening phase during which
some very bright outbursts were observed.
The first event  lasted about 10 days in August 2009 (Bastieri, 2009) and  was characterized by fast rise and decay times ($< 1$ day) and a relatively
stable `plateau'. In contrast the second and third otbursts were  sharply peaked from September 15 to 17 and from September 20 to 23, 2009 (Hill, 2009), when \tct reached a peak photon flux above $10^{-5}$ \latflux.
Before the very recent (December 2009) flare observed in 3C 454.3 (Escande \& Tanaka, 2009;
Stirani et al., 2009), it was, therefore, the brightest extragalactic source, non Gamma-Ray Burst, observed by {\it Fermi}. EGRET detected only a few blazars (e.g. 3C 279 and  PKS 1622$-$297) with a flux above $10^{-5}$ \latflux.

Thanks to the high flux is was possible to obtain light curves with a good S/N
ratio with a time binning of only six hours (corresponding to two scans of
sky) and therefore  we can describe the evolution of this outburst
with a level of detail never reached before.
In this letter we present the results of the LAT observations.

\section{Observations with the Large Area Telescope }

The {\it Fermi}-LAT is a pair-conversion $\gamma$-ray telescope sensitive to
photon energies from 20~\MeV~ to $>300$~\GeV.
It is made of a tracker (composed of two sections, front and back, with
different capabilities), a calorimeter, and an anticoincidence system to
reject the charged-particle background.
The LAT has a large peak effective area ($\sim8000$ cm$^2$ for 1 \GeV~
photons in the event class considered here), viewing $\sim$2.4 sr of the full
sky with excellent angular resolution (68\% containment radius better than
$\sim1^{\circ}$ at $E = 1$ \GeV).
A full description of the LAT instrument and its performance are reported in Atwood et al. (2009).
During the first year, the telescope operated in sky-survey mode, observing
the whole sky every 3 hours.

\begin{figure}
\plotone{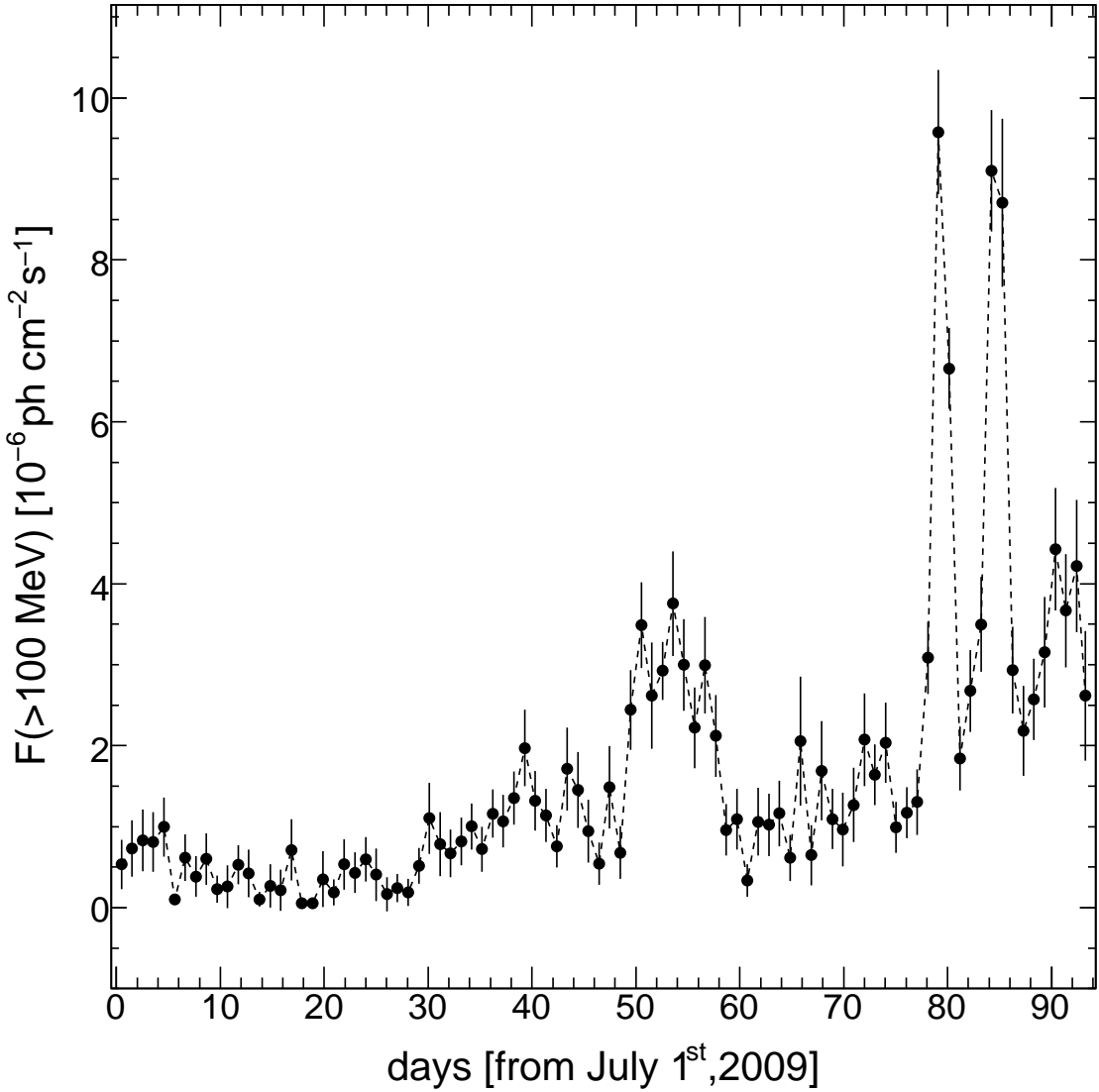}
\caption{The 1-day gamma-ray light curve of \tct in the period from July 1 (MJD=55013) to September 30, 2009 (MJD=55104).}
\label{fig:fig1}
\end{figure}

Only photons in ``diffuse" class with energies greater than 100 MeV were
considered in this analysis.
We kept only events with reconstructed zenith angle $<105^{\circ}$ in order
to reduce the bright $\gamma$-ray albedo from the Earth.
In addition, we excluded the time intervals when the rocking angle was more
than 52$^{\circ}$ and when the {\it Fermi} satellite was within the South
Atlantic Anomaly.
The standard {\it Fermi-LAT ScienceTools} software package\footnote{http://fermi.gsfc.nasa.gov/ssc/data/analysis/documentation/Cicerone/} (version v9r15p6 ) was used with the P6 V3  set of instrument response functions. This version of the Science Tools takes into account the correction for the residual acceptance variation with count rate.
Light curves were produced in 1-day (see Figure 1) bins over the period July-September 2009,
and in 12 and 6-hour (see Figure 2) bins in September 2009.
For each time bin, the flux, photon index and test statistic of \tct  were determined, in the energy range 0.1-100 GeV. We analyzed a Region of Interest (RoI) of 12$^{\circ}$ in radius, centered at
the position of the $\gamma$-ray source associated with \tct, using the maximum-likelihood algorithm implemented in {\it gtlike}.  In the RoI analysis the sources were modeled as simple power-law  ($F = K E^{-\Gamma}$).  The source model parameters are free for all point sources within $5^{\circ}$ of \tct (extracted from the LAT 1FGL catalog\footnote{http://fermi.gsfc.nasa.gov/ssc/data/access/lat/1yr\_catalog/}, see Abdo et al. 2010a).  In the RoI model, we also included all sources from $5^{\circ}$ to $17^{\circ}$ of \tc, with their model parameters fixed to their catalog values.

The Galactic diffuse background model included in the RoI analysis is the currently recommended version (gll\_iem\_v02), publicly released through the {\it Fermi} Science Support
Center. The isotropic background (including the $\gamma$-ray diffuse and residual
instrumental backgrounds) model was derived from an overall fit of the diffuse
component over the $|b|>30\arcdeg$ sky. A detailed documentation of the Galactic diffuse model and corresponding isotropic spectrum is available from the {\it Fermi} Science Support Center\footnote{http://fermi.gsfc.nasa.gov/ssc/data/access/lat/BackgroundModels.html}.

All errors reported in the figures or quoted in the text are 1-$\sigma$ statistical errors. The estimated systematic uncertainty on the flux is 10\% at 100 MeV, 5\% at 500 MeV and 20\% at 10 GeV.

\section{Time evolution of the outbursts}

The light curve of the very large outburst for photon energies above 0.1 \GeV~
and with a time binning of 6 hours is plotted in the upper panel of Fig. 2.
The outburst's structure consists of two main flares, each one having a total
duration of about a couple of days and peak flux of $\sim$1.2\latfluxv.
The first flare shows a nice smooth time profile, suggesting that it is well
resolved; the second flare, after a first short peak, shows a series of high
bins that can be indication either of substructures or of a long decay tail.

The profile of the first flare is clearly asymmetric with a fast rise of the
duration of about 12 hours, and a longer decay of about 36 hours.
We modelled its evolution by means  of the law of the following formula:
\begin{equation}
F(t)=F_{o}+K (1+t/T_1)^{\alpha}(1-t/T_2)^{\beta}
\end{equation}
that is derived from the statistical Beta distribution (see, for instance Johnson, Kotz, \& Balakrishnan 1994) and is also known as a Pearson curve of Type I (Smart 1958).
It is defined in the interval $(-T_1, T_2)$ (a shift of the time origin is necessary
for the fitting to data) and  $F(-T_1)=F(T_2)=F_o$ at the extremes of that interval, then it is well suitable to model burst of finite durations.
This function can be used to model the time evolution of flares, having
either convex or concave rising and decaying profiles. It is more general
than the one applied in Abdo et al. (2010c), which presented and analyzed the
gamma-ray light curves of the 106 high-confidence Fermi LAT Bright AGN Sample
(Abdo et al. 2009) obtained during the first 11 months of the {\it Fermi} survey.
The flares of the 10 brightest blazars were modelled by a law, defined as the
inverse of the sum of two exponentials, to quantify the symmetry and duration
of individual episodes.
The total duration of the flare is given by $T_1 + T_2$, while we can easily measure
its asymmetry by evaluating the difference between the decay and the rise
time divided by their sum, which gives:
\begin{equation}
\xi = \frac{\beta - \alpha}{\alpha + \beta} ~~~~~.
\end{equation}

For the first burst we find $T_1=18$ hours,  $T_2=60$ hours,$\alpha=0.5$, $\beta=3.2$ and $\xi=0.73$.
The complex structure of the second flare make  the fitting of Eq. (1) very hard,
however, considering only the first five data points we obtained a reasonable fit with  $T_1=12$ hours and  $T_2=33$ hours and using the same values of $\alpha$ and $\beta$ (see Fig. 2). The resulting $\chi^2$ are smaller than unity in both cases.
This result differs from those of the other six and much longer bursts observed
by the LAT to July 2009 (Abdo et al. 2010c).
These flares had durations in the range from 8 to 58 days and were generally
nearly symmetric, all they had negative values $\xi$, usually in the interval
($-0.38, -0.22$) with the only exception being the longest one for which $\xi$
was $-0.72$.
Although the function used in the flare analysis of \tct presented by Abdo et al. (2010c) was not that of Eq. (1),  the quantity $\xi$ was the same and depends only on the time elapsed between the peak and the starting and ending points of the flare. The highest brightness level reached in those flares, however, was much lower
than that of the September 2009 outbursts: the typical flux of \tct was around
3\latfluxviiiii~ with peak values during the flares of 2.5\latfluxvi.
Comparable fluxes were measured by the GRID instrument, on board the AGILE satellite, from
December 2007 to January 2008 (Pacciani et al. 2009).

It is interesting to evaluate the rising and decay flux rates of the two
flares.
The rises were both very short, and the estimates of the corresponding
flux rates at about half of the rising branch are 1.3\latfluxviiii and
2.0\latfluxviiii for the first and second flare, respectively.
The decays evolved on longer time scale with flux rates of
0.3\latfluxviiii and 0.5\latfluxviiii.

The very fast variability, however, does not correspond to an increase
of the time-integrated energy output from the source, at least when they
are compared with that of August.
In fact the fluence of this flare, estimated by the integration of the best fit
model over the duration of the event, is 2.06 \fluence, while those of the other
and higher two flares are 1.59 and 1.05 \fluence.
The last value is that of the main peak and increases by about 0.5 \fluence when the
three last bins are included.

A comparison of the power density spectrum (PDS) of the 90 day flaring
episode in Fig. 1 with one computed for the first 11 months of data,
did not show any substantial difference in shape, although the
flare PDS had slightly more power at frequencies $>$0.1 days$^{-1}$.
The fractional variability (rms/flux) for frequencies above 0.01 days$^{-1}$
in the PDS, was however, three times higher during the flaring episode
than during the initial 11 months (0.99 and 0.32 respectivelly). In other words, the source was more variable during the flaring period of July-September than in the preceding 11 months.


\begin{figure}
\plotone{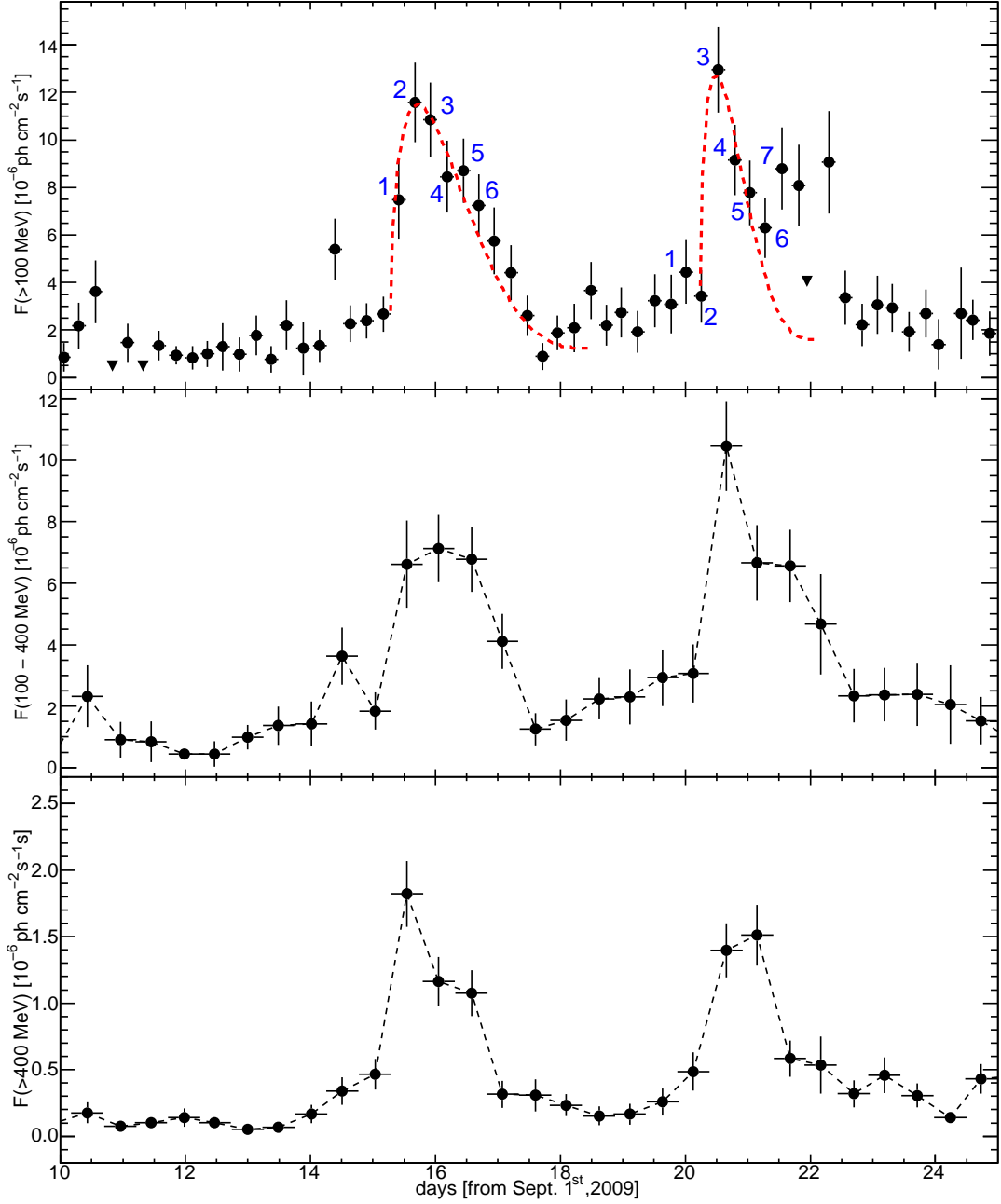}
\caption{The upper panel shows the 6-hour light curve of \tct during the large outbursts in September 2009. The points marked with triangles indicate upper limits. The middle and lower panels show 12-hour light curves of 3C 273 during the large outburst in September 2009 in the energy range 0.1-- 0.4 GeV and at energies higher than 0.4 GeV, respectively.}
\label{fig:fig2}
\end{figure}

\section{Spectral behavior}

\begin{figure}
\centering
\includegraphics[height=100mm]{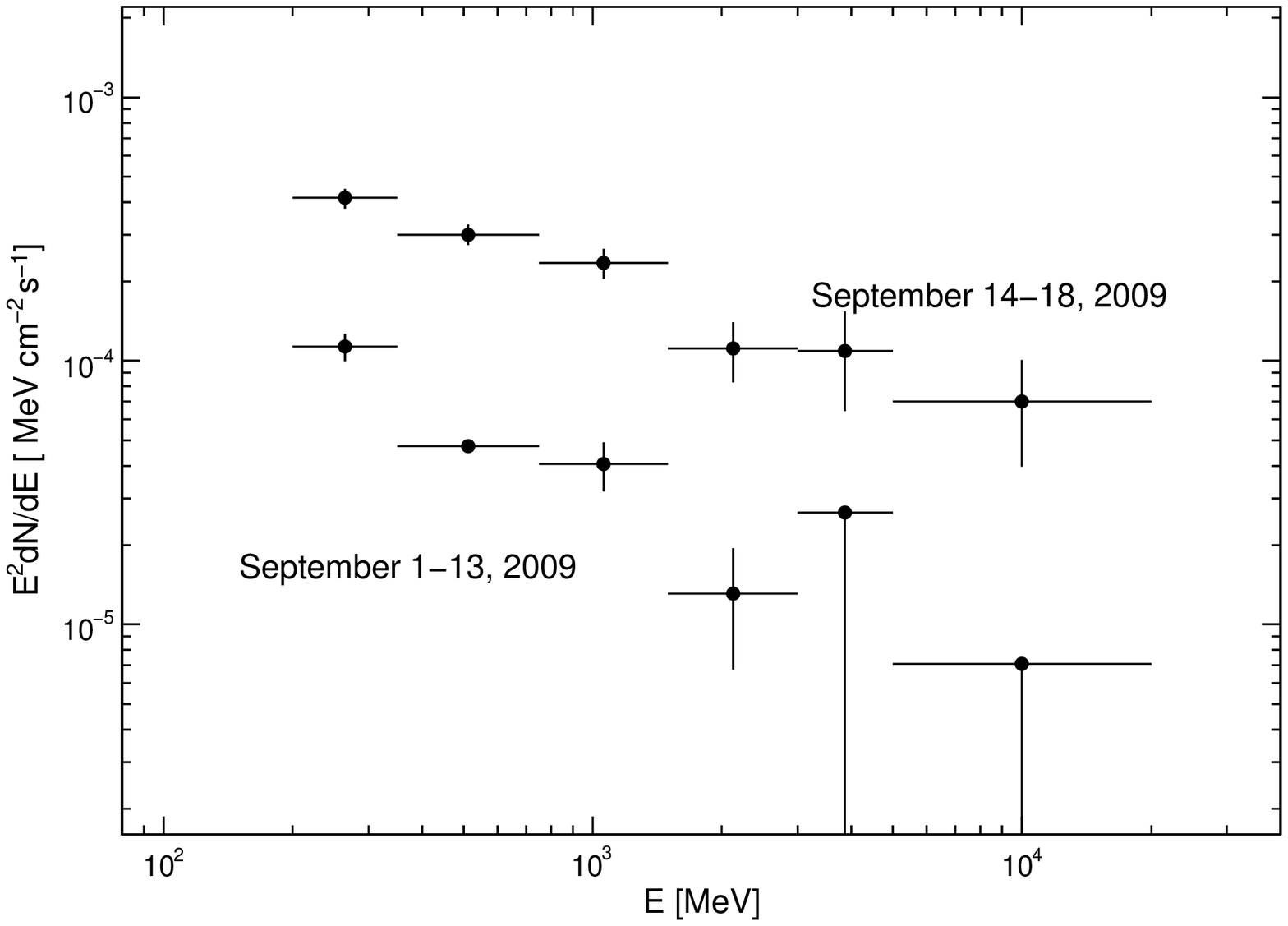}\\
\includegraphics[height=100mm]{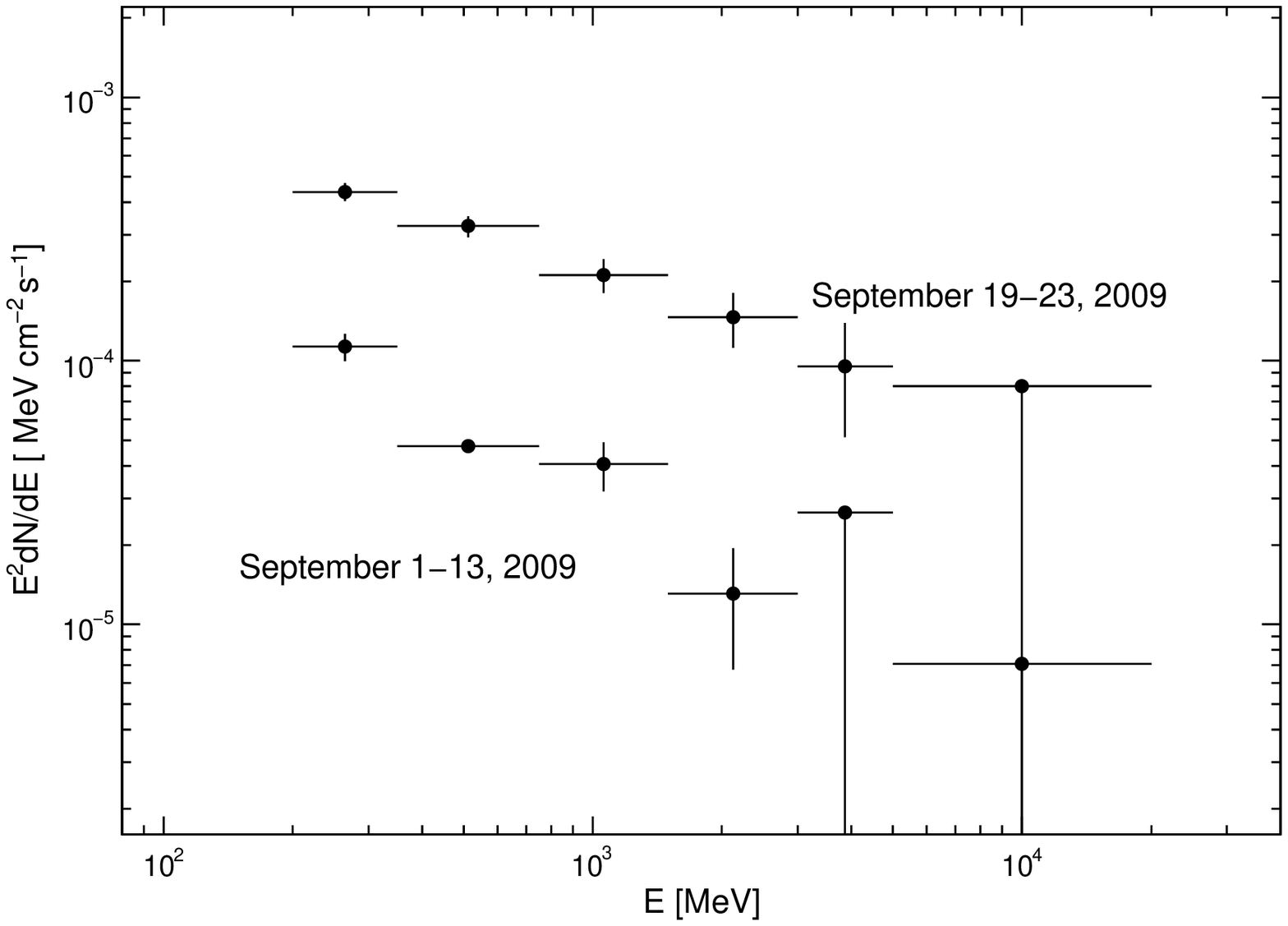}
\caption{The Spectral Energy Distributions of 3C~273 in the quiescent phase
and during the two September outbursts.}
\label{fig:fig3}
\end{figure}

Figure \ref{fig:fig3} shows the spectral energy distribution, in the band 100 MeV -- 20 GeV,
of \tct during the September outbursts.
The spectra were obtained fitting a power law model over seven
energy bins, with the spectral index kept fixed at the values obtained by the
fit in the entire energy range over the same time interval.

We also computed the spectra of \tct integrating over the two weeks from September
1 to 13, during the quiescent phase preceeding the September events, and found that
it is well described by a power law with photon index $\Gamma$=2.86$\pm$0.02
(see Figure \ref{fig:fig3}).
The spectra of the two major outbursts integrated over the time intervals
from 14 to 18 September and from 19 to 22 September 2009, showed
flatter slopes $\Gamma$= 2.43$\pm$0.01 and 2.52$\pm$0.01,
respectively, confirming that the average spectrum during the oubursts
was significantly but moderately harder than in the quiescent state.
As reported in Abdo et al.(2010d)  the spectrum of 3C273 integrated over the 6-month showed a significant break around 1.5 GeV.  Our spectra of the source integrated over shorter periods  of time do not provide indication for a curvature in the spectra.

It is to note that so far the spectral change with respect to flux has been observed to be very moderate in other FSRQs (e.g. 3C 454.3, see Abdo et al. 2010d), even during large flares.
An interesting indication of a possible spectral evolution during the major
bursts of \tct can be obtained by the light curves in two energy bands.
The central and lower panel of Figure 2 shows the outburst evolution
in the energy ranges (0.1 -- 0.4) \GeV~ and above 0.4 \GeV~with a time
bin width of 12-hour.
Despite the large differences in the photon fluxes, we see that
the profile of the peaks are rather similar with an indication that above 0.4
\GeV~ peaks are sharper and with shorter decay times.

Time resolved spectral indices show possible spectral changes during the two
major bursts as illustrated by the $\Gamma$ vs Flux plots in Figure 4.
The start and end ranges of the data used to demarcate the two flares were
defined as the time intervals containing the 90\% of the total flux,
starting when the 10\% level was measured.
A ``counterclockwise" loop  is observed during the first flare.
The spectrum becomes steeper in the declining phase and harder in
the brightening phase.
During the second flare the spectral index changes following a path that is essentially  ``clockwise" . In this case the brightening phase was very rapid and the spectral
index remained constant, after the flare peak the spectrum becomes harder and
then softer again.
A clockwise loop can indicate that during the first flare the flux started to increase
at low energy and then propagate to high energy, the inverse could have happened
during the second flare.

\begin{figure}
\centering
\includegraphics[height=100mm]{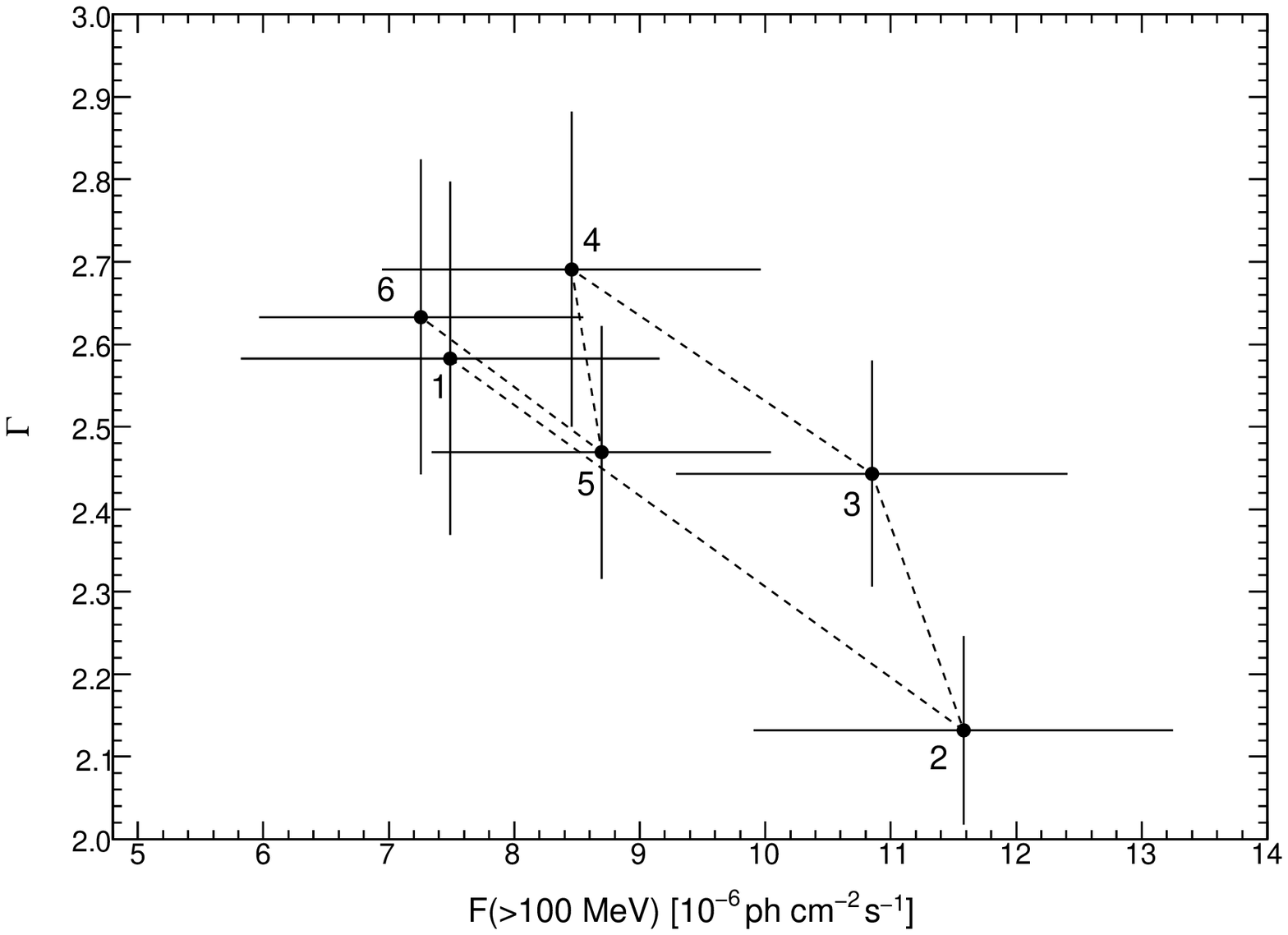}\\
\includegraphics[height=100mm]{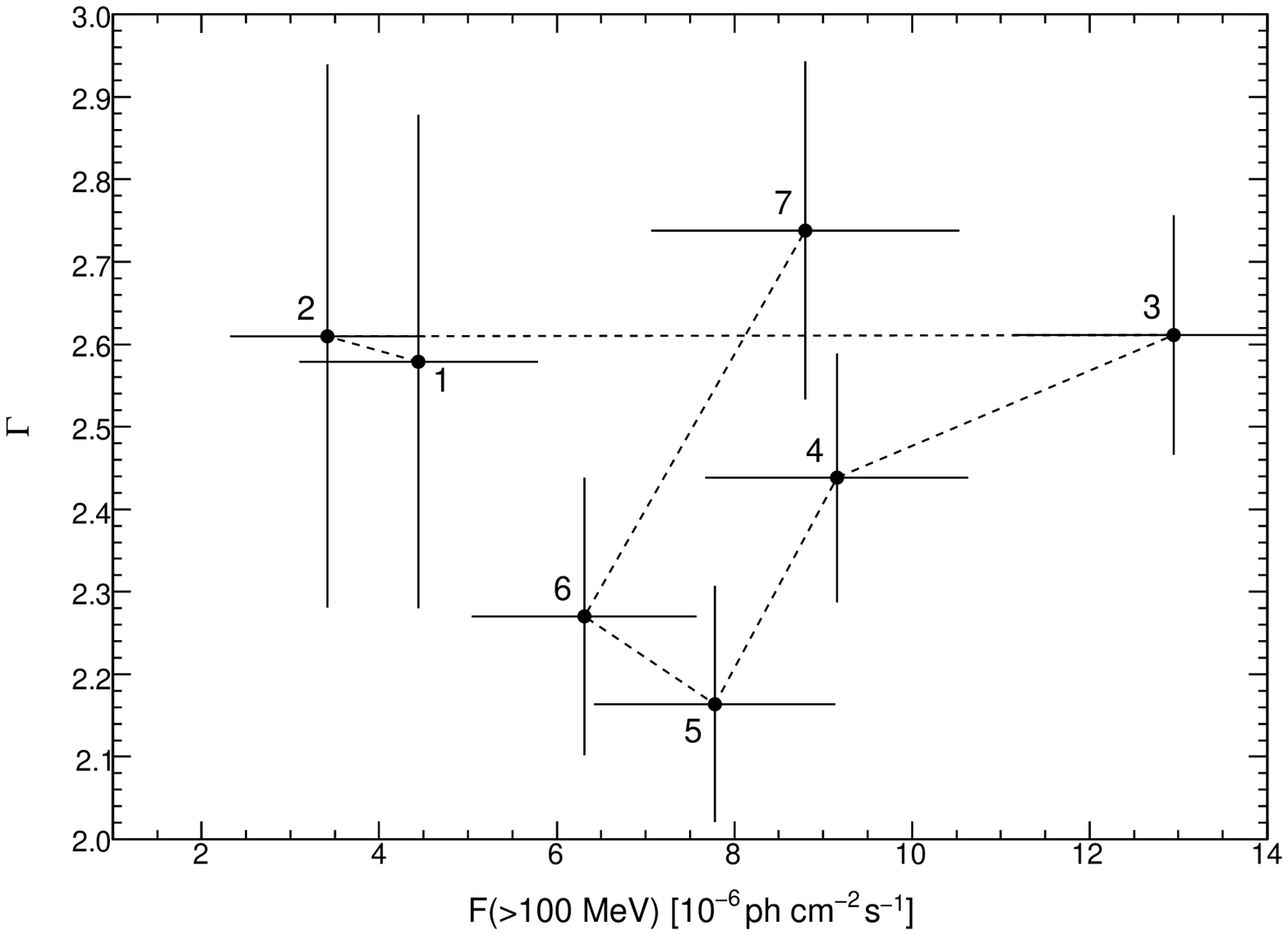}
\caption{This figure show the variation of $\Gamma$ during the large flares observed
in September 15-16, 2009 (upper panel) and September 19-21, 2009 (lower panel). In the upper panel the point having the label (1) correspond  to September 15.30, 2009,  the other points corresponds to successive  6-hour time bins (see also Figure \ref{fig:fig2}, upper panel). In the lower panel  the point having the label (1) correspond  to September 19.88, 2009, the other points corresponds to successive  6-hour time bins  (see also Figure \ref{fig:fig2}, upper panel).}
\label{fig:fig4}
\end{figure}

\section{Discussion}

The large outburst of \tct observed by {\it Fermi} in September 2009 revealed
that blazars can reach very high brightness levels for quite short short time intervals of 1-10 days.

The fluences of these events, however, are comparable to that of longer
but less intense flares.
We measured significant variations of the flux over
time scales as short as six hours, which occurred with only a mild change of the spectral shape.

In contrast to previous low intensity flares (Abdo et al., 2010c), the very strong events of September
have light curves characterized by decay times longer than the rise time.
Light curves at different energies provide some indication, in particular
for the second flare, that the decay rate above 0.4 \GeV~ is shorter than below.
The most natural interpretation of this difference is that  it is
due to the radiative cooling of the high energy electrons responsible for the $\gamma$-ray emission.
We can therefore to estimate that this time, as observed in the Earth's frame,
is of the order of 0.5--1 day.
Such short lifetimes explain why longer flares are generally symmetric.
In fact, it is possible that they are structured in a series of short subflares
with typical durations of two-three days, much shorter than the total flare
length.
The apparent rise and decay times of long events are then likely due to the
superposition of quite shorter flares and any information on the radiative lifetimes is practically lost.

In September 2009 the apparent position of \tct was rather close to the Sun,
making it was practically impossible to perform observations in other frequency ranges.
Therefore we do not have data to study the evolution of the broad band
Spectral Energy Distribution.
The typical SEDs of FSRQs (Abdo et al. 2010b) peak at rather low $\gamma$-ray
energies, around 0.1 GeV and frequently lower.
In the case of \tc, which was the most significant AGN detected by COMPTEL in the
1-30 MeV range, the peak of the inverse Compton component was in the range 1--10
MeV (Collmar et al 2000).
Our data show that the photon index was always steeper than 2, and this
indicates that the peak energy remained below 100 MeV also during the flares.


$\gamma$-ray activity has been found to be related to changes of the blazars'
radio structure observed with VLBI.
In this respect \tct is one of the most interesting sources because of the
flux and of the low redshift that allows a fine spatial resolution.
Jorstad et al. (2001), on the basis of a large data set at 22 and 43 GHz,
found an association between the ejection of superluminal radio knots and high
states of $\gamma$-ray luminosity in ten blazars in EGRET observations,
including 3C~273. They concluded that both the radio and high energy events are originating
from the same shocked region of a relativistic jet.
Similar correspondences were already reported for the two much more distant
FSRQs S5 0836+710 (Otterbein et al. 1998) and PKS 0528+134 (Britzen et al. 1999).

More recent results on the MOJAVE VLBA sample and Fermi-LAT observations (Lister
et al. 2009, Savolainen et al. 2009a) provided evidence that $\gamma$-ray loud
blazars have a Doppler factor higher than non LAT-detected sources.
A detailed plot of the kinematics of the various components in the radio jet
of \tct can be found in Lister et al. (2009b): the resulting mean superluminal
velocity is $\beta_{app}$=13.4 with an estimated Doppler factor $\delta$=16.8.

It will be very interesting to verify if the exceptional outbursts of September
2009 will or will not be associated with the ejection of new superluminal knots,
possibly with even higher velocity and Doppler factor.
Moreover, the discovery of a possible connection between the peak intensity and
rise time of the $\gamma$-ray outbursts with the VLBI parameters can be very
useful to constrain the modelling and the energetics of perturbations in the jet.

\section{Acknowledgments}
\acknowledgments

The \textit{Fermi} LAT Collaboration acknowledges generous ongoing support
from a number of agencies and institutes that have supported both the
development and the operation of the LAT as well as scientific data analysis.
These include the National Aeronautics and Space Administration and the
Department of Energy in the United States, the Commissariat \`a l'Energie Atomique
and the Centre National de la Recherche Scientifique / Institut National de Physique
Nucl\'eaire et de Physique des Particules in France, the Agenzia Spaziale Italiana
and the Istituto Nazionale di Fisica Nucleare in Italy, the Ministry of Education,
Culture, Sports, Science and Technology (MEXT), High Energy Accelerator Research
Organization (KEK) and Japan Aerospace Exploration Agency (JAXA) in Japan, and
the K.~A.~Wallenberg Foundation, the Swedish Research Council and the
Swedish National Space Board in Sweden.

Additional support for science analysis during the operations phase is gratefully
acknowledged from the Istituto Nazionale di Astrofisica in Italy and the Centre National d'\'Etudes Spatiales.

{\it Facilities:} \facility{{\it Fermi} LAT}.




\end{document}